\documentclass[runningheads]{llncs}
\usepackage{caption}
\usepackage{subcaption}
\usepackage{booktabs}
\usepackage{multirow}
\usepackage{graphicx}
\usepackage{hyperref}
\usepackage[subtle,margins=normal,leading=normal]{savetrees}
\usepackage{pifont}
\newcommand{\cmark}{\ding{51}}%
\newcommand{\xmark}{\ding{55}}%
\usepackage[T1]{fontenc}
\usepackage{graphicx}

\begin{document}
\title{Adult Glioma Segmentation in Sub-Saharan Africa using Transfer Learning on Stratified Finetuning Data}
\titlerunning{SSA: Adult Glioma Segmentation}
%
\author{Abhijeet Parida\inst{1,2}*, Daniel Capell\'{a}n-Mart\'{i}n\inst{1,2}*, Zhifan Jiang\inst{1}*, \\ Austin Tapp\inst{1}, Xinyang Liu\inst{1}, Syed Muhammad Anwar\inst{1,3}  ,\\ Mar\'{i}a J. Ledesma-Carbayo\inst{2} and Marius George Linguraru\inst{1,3}}

\authorrunning{A. Parida, D. Capell\'{a}n-Mart\'{i}n, Z. Jiang et al.}
%
\institute{Sheikh Zayed Institute for Pediatric Surgical Innovation, \\ Children’s National Hospital, Washington, DC, USA \and Biomedical Image Technologies, ETSI Telecomunicaci\'{o}n, \\ Universidad Polit\'{e}cnica de Madrid \& CIBER-BBN, ISCIII, Madrid, Spain \and School of Medicine and Health Sciences, \\George Washington University, Washington, DC, USA}

\maketitle              
\begin{abstract}
Gliomas, a kind of brain tumor characterized by high mortality, present substantial diagnostic challenges in low- and middle-income countries, particularly in Sub-Saharan Africa. This paper introduces a novel approach to glioma segmentation using transfer learning to address challenges in resource-limited regions with minimal and low-quality MRI data. We leverage pre-trained deep learning models, nnU-Net and MedNeXt, and apply a stratified fine-tuning strategy using the BraTS2023-Adult-Glioma and BraTS-Africa datasets. Our method exploits radiomic analysis to create stratified training folds, model training on a large brain tumor dataset, and transfer learning to the Sub-Saharan context. A weighted model ensembling strategy and adaptive post-processing are employed to enhance segmentation accuracy. The evaluation of our proposed method on unseen validation cases on the BraTS-Africa 2024 task resulted in lesion-wise mean Dice scores of 0.870, 0.865, and 0.926, for enhancing tumor, tumor core, and whole tumor regions and was ranked first for the challenge. Our approach highlights the ability of integrated machine-learning techniques to bridge the gap between the medical imaging capabilities of resource-limited countries and established developed regions. By tailoring our methods to a target population's specific needs and constraints, we aim to enhance diagnostic capabilities in isolated environments. Our findings underscore the importance of approaches like local data integration and stratification refinement to address healthcare disparities, ensure practical applicability, and enhance impact. 
\\

* These authors contributed equally.

\end{abstract}
\section{Introduction}

Gliomas, a type of brain tumor known for their high mortality rates, exhibit poor survival outcomes, with significant disparities between high- and low-income countries \cite{di2024economic}. While some progress has been made in reducing mortality rates in high-income countries like the United States, low- and middle-income countries, notably those in Sub-Saharan Africa, continue to face increasing glioma death rates \cite{lin2021trends}. Death rate disparity is primarily due to delayed diagnosis, comorbidities such as HIV, and inadequate healthcare infrastructure \cite{yevudza2024neuro}. Machine learning (ML) offers promise in narrowing this survival gap by enhancing early detection, treatment precision, and outcome prediction through improved MRI-based tumor segmentation. However, adapting advanced ML methods to regions with limited medical resources and poorer MRI quality remains a significant challenge \cite{brats-ssa2023}.

Healthcare systems in Sub-Saharan Africa often struggle with limited access to high-quality imaging equipment, a scarcity of annotated medical data, and inadequate computational resources \cite{vibbi2024poor}. These constraints hinder the development and deployment of robust glioma segmentation models tailored to the local population \cite{yevudza2024neuro}. Therefore, innovative approaches are urgently needed to overcome these barriers and provide accurate and reliable segmentation results under such conditions.

The Brain Tumor Segmentation (BraTS) challenge, held in conjunction with the International Conference on Medical Image Computing and Computer-Assisted Intervention (MICCAI) since 2012, has established a benchmark dataset for the segmentation of adult brain gliomas \cite{brats2021,bakas1,bakas2,bakas3,brats2015}. The BraTS 2024 cluster of challenges \cite{medperf} has reintroduced the benchmark for adult brain gliomas \cite{GLIarxiv2024}, specifically targeting the Sub-Saharan African (SSA) patient population \cite{brats-ssa2023}. The MICCAI-CAMERA-Lacuna Fund BraTS-Africa 2024 Challenge \cite{brats-ssa2023} provides the largest annotated publicly available retrospective cohort of pre-operative glioma in adult Africans, including both low-grade glioma and glioblastoma/high-grade glioma. In this context, we propose a segmentation technique for benchmarking the SSA task, which involves a small dataset with low-quality MRI acquisition.

In this context, transfer learning has emerged as a powerful technique in medical imaging, offering a solution to the challenges posed by limited data availability and/or quality \cite{yu2022transfer,atasever2023comprehensive}. By leveraging pre-trained models on large, diverse datasets, transfer learning enables adapting these models to specific tasks with significantly fewer training samples \cite{dicom,cxr}. This approach mitigates issues of data scarcity and enhances model performance by incorporating learned features from extensive, high-quality datasets. Moreover, in the medical domain, supervised pre-training is particularly beneficial to improve learning and address the problem of catastrophic forgetting \cite{lee2024supervised,liao2022muscle}.

In this work, we develop an ensemble approach involving two state-of-the-art deep learning models. These models are specifically trained for glioma segmentation under low-data and low-quality conditions. We implement a stratified fine-tuning strategy tailored to the Sub-Saharan context and thoroughly evaluate the model's effectiveness in accurately segmenting gliomas in local MRI scans. Through this work, we aim to demonstrate the potential of advanced machine learning techniques to bridge the gap in medical imaging capabilities between resource-limited regions and more developed healthcare systems.

\begin{figure}[htbp]
  \centering
  \includegraphics[width=0.99\textwidth]{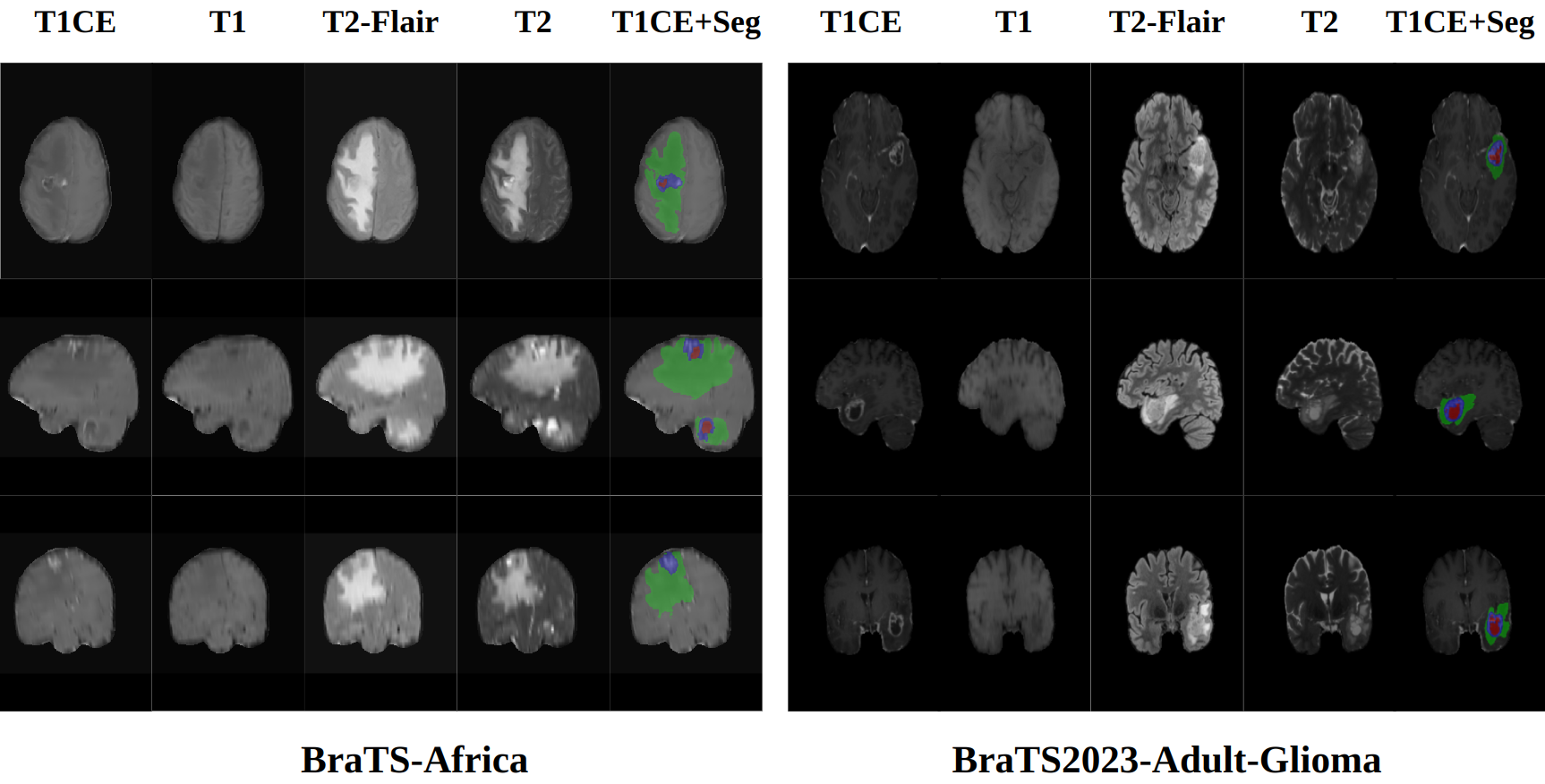}
  \caption{\textbf{Training examples} showing the differences between the BraTS-Adult-Glioma and the BraTS2023-Africa datasets. }
  \label{fig:datasets}
\end{figure}
\section{Data description}

For our model training, we utilized the BraTS2023-Adult-Glioma \cite{brats2021} and BraTS-Africa datasets \cite{brats-ssa2023}. These datasets were collected during routine multi-parametric MRI (mpMRI) clinical scans, acquired as part of standard clinical care from multiple institutions using conventional brain tumor imaging protocols. The mpMRI data included pre- and post-gadolinium T1-weighted (T1 and T1CE), T2-weighted (T2), and T2-weighted fluid-attenuated inversion recovery (T2-FLAIR) MRI scans.

\begin{table}[h]
\caption{\textbf{Dataset description} of the datasets used for training and validation for the challenge and in which stage they were used. The datasets have an additional unknown number of testing sets.}
\centering
\resizebox{0.8\columnwidth}{!}{%
\begin{tabular}{l|cc|cc}
\hline
                   & Training & Validation & Pretraining & Finetuning \\ \hline
BraTS2023-Adult-Glioma &     1251     & -NA-       &      \cmark       & \xmark           \\
BraTS-Africa       & 60       & 35         &      \xmark       &    \cmark        \\ \hline
\end{tabular}%
}
\vspace{0.2cm}

\label{tab:data}
\end{table}

The mpMRI scans underwent standardized pre-processing according to the BraTS challenge guidelines \cite{10.3389/fnins.2020.00125}. This pre-processing included co-registration to the SRI Atlas \cite{Rohlfing2010}, resampling to an isotropic resolution of 1$mm^3$, and skull-stripping. The ground truth annotations for the tumor sub-regions—Peritumoral Edema (ED), Enhancing Tumor (ET), and Necrotic Core (NC)—for each case were approved by expert neuroradiologists.

Fig. \ref{fig:datasets} and Tab. \ref{tab:data} show samples of the training examples used for the task, providing a visual and quantitative overview of the data utilized in our model training.

\section{Methods}
Our methodology for generalizable brain tumor segmentation on low-quality and limited data leverages commonly used segmentation architectures with transfer learning. It involves radiomic analysis to create stratified training folds, model training on a large brain tumor dataset, transferring learning to the SSA dataset, weighted model ensembling, and postprocessing.
\begin{figure}[htbp]
  \includegraphics[width=\textwidth]{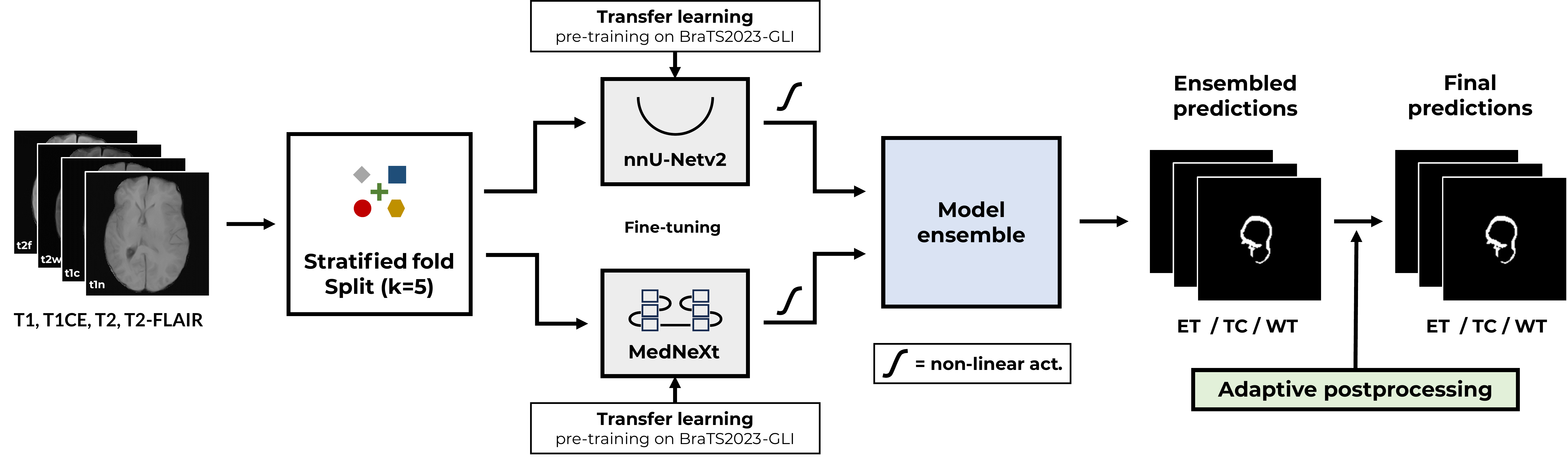}
  \caption{\textbf{Proposed method:} Unsupervised stratified fold split, fine-tuning of pre-trained models on BraTS2023-GLI, model ensembling and adaptive post-processing. Outputs are obtained from two state-of-the-art deep learning models. These outputs are subjected to nonlinear activation functions and ensembling strategies. Finally, the ensembled predictions are subjected to a specifically tailored adaptive post-processing step.}
  \label{fig:pipeline}
\end{figure}

\subsection{Stratified Fold Creation}\label{section:fold}
Deep learning tasks often use random sampling to create folds, but this can result in unbalanced learning because not all data types are equally represented in each fold. To address this issue, we created stratified folds for our deep learning models, ensuring a good representation of the radiomic features of gliomas in each fold.

Following the approach in \cite{jiang_automatic_2023}, we computed 14 shape-based and 93 intensity-based radiomic features on the largest lesion area (the whole tumor, or WT) for each MRI sequence. We used principal component analysis (PCA) to select the most relevant features, which explained 99\% of the variance, resulting in 9 features. These features were then used in a k-means clustering algorithm to group lesions into different clusters (subtypes) of tumors. The optimal number of clusters was determined using grid search and silhouette analysis.

The k-means algorithm was trained on the training set with the corresponding ground-truth WT for tumor subtype analysis during preprocessing and on the cross-validated predicted WT during post-processing. Equal samples from each of the k-means clusters were then used to create our stratified folds.

\subsection{Deep Learning Models and Transfer Learning Setup}

\subsubsection{nnU-Net}
The nnU-Net is a self-configuring deep learning framework for semantic segmentation based on the U-Net architecture \cite{unet}. It automatically adjusts its internal configurations according to the specific imaging modality and unique attributes of each dataset \cite{nnunet}. The self-configuration process leads to improved segmentation performance and generalization compared to other state-of-the-art methods for biomedical image segmentation.

For \textbf{pretraining}, we trained a full-resolution 3D nnU-Net (v2) model using a stratified five-fold from the \textit{BraTS2023-Adult-Glioma} dataset. The model predicted three channels corresponding to the three tumor sub-regions (enhancing tumor, tumor core, and whole tumor). The input images were preprocessed using zero mean unit variance normalization and divided into patches of 128x128x128 voxels. We used region-based training to favor larger patches while staying within the GPU's capacity \cite{nnunet}.

The loss function combined Dice loss and cross-entropy loss, optimized with the stochastic gradient descent (SGD) optimizer using Nesterov momentum. We set an initial learning rate of 0.01, a momentum of 0.99, and a weight decay of 3e-05. Each of the five folds was trained for 200 epochs on an NVIDIA A100 (40 GB) GPU. During inference, images were predicted using a sliding window approach, with the window size matching the patch size used during training.

For \textbf{transfer learning}, we fine-tuned the pretrained full-resolution 3D nnU-Net (v2) model on a stratified five-fold from the \textit{BraTS-Africa} dataset, using the same configuration as the pretraining. The nnU-Net implementation is available in an open-source repository: \href{https://github.com/MIC-DKFZ/nnUNet}{https://github.com/MIC-DKFZ/nnUNet}.

\subsubsection{MedNext}
MedNext \cite{mednext} combines convolutional neural networks and attention mechanisms for medical image analysis. It integrates convolutional layers for feature extraction with attention modules that enhance the model’s focus on relevant regions within the images \cite{mednext}. Using strategies from the 3D nnU-Net (v2) \cite{nnunet}, the framework autonomously adjusts its internal configurations for better performance.

Like 3D nnU-Net, MedNext was trained in a label-respective manner for each task. For \textbf{pretraining} on the stratified fold of \textit{BraTS2023-Adult-Glioma} dataset, we trained MedNeXt-M (k=3, 17.6M parameters, 248 GFlops) with a class-weighted loss function combining Dice loss and cross-entropy loss. The loss function was optimized with the SGD optimizer using Nesterov momentum, with an initial learning rate of 0.01, momentum of 0.99, and weight decay of 3e-05. Each of the five folds was trained for 200 epochs on an NVIDIA A100 (40 GB) GPU. During inference, images were predicted using a sliding window approach, with the window size matching the patch size used during training.

For \textbf{transfer learning}, we fine-tuned the pre-trained MedNext model on a stratified five-fold from the \textit{BraTS-Africa} dataset, using the same configuration as the pretraining. The MedNext implementation is available in an open-source repository: \href{https://github.com/MIC-DKFZ/MedNeXt}{https://github.com/MIC-DKFZ/MedNeXt}.

\subsection{Weighted Model Ensembling}

We propose to use a model ensembling strategy to enhance the accuracy and robustness of the segmentation outcome \cite{brats2023sub}. This approach (see Fig. \ref{fig:pipeline}) involves harnessing the complementary strengths of the two models described, nnU-Net and MedNeXt, to collectively address the task of pixel classification. Each model is trained on stratified five folds that provide a similar representation, so we assign equal weight to each fold within the model. However, when combining the models, we use a weighted ensemble represented by the equation below:

\begin{equation}
    X_{ens} = w_n * nnU\-Net(X) + w_m * MedNeXt(X)
    \label{eqn:1}
\end{equation}

\subsection{Adaptive Post-processing}\label{sec:adapt}

Adaptive post-processing involves choosing post-processing parameters based on the ensemble prediction for each new sample \cite{brats-goat}. We calculate 14 shape-based and 93 intensity-based radiomic features on the predicted whole tumor (WT) for this. Using PCA analysis (see section \ref{section:fold}) on the WT predictions, we identified 8 clusters for post-processing.

The BraTS challenge evaluates segmentation models at the lesion level rather than across entire tumor regions. To refine our predictions, we performed a grid search within each cluster to find optimal thresholds for removing small, isolated areas likely to be noise, as identified from cross-validation data.

Additionally, we conducted a further threshold search on the refined segmentation maps to adjust the ET label based on the ET/WT ratio \cite{brats2023sub}. In this step, if the ET/WT ratio was below a certain threshold, we redefined the ET label to NCR. This adjustment was made for all 8 clusters using cross-validated results.

\section{Results}
\subsection{Evaluation Criteria}
The evaluation of model predictions on the validation set utilized the BraTS pipeline within the Synapse platform. Assessments were conducted for distinct tumor regions including the whole tumor (WT), tumor core (TC), and enhancing tumor (ET). Evaluation metrics included the Dice score for lesion-wise overlap between predicted segmentation and ground truth, and the 95th percentile lesion-wise Hausdorff distance to measure segmentation deviation from the ground truth.

\begin{figure}[htbp]
    \centering
  \includegraphics[width=0.7\textwidth]{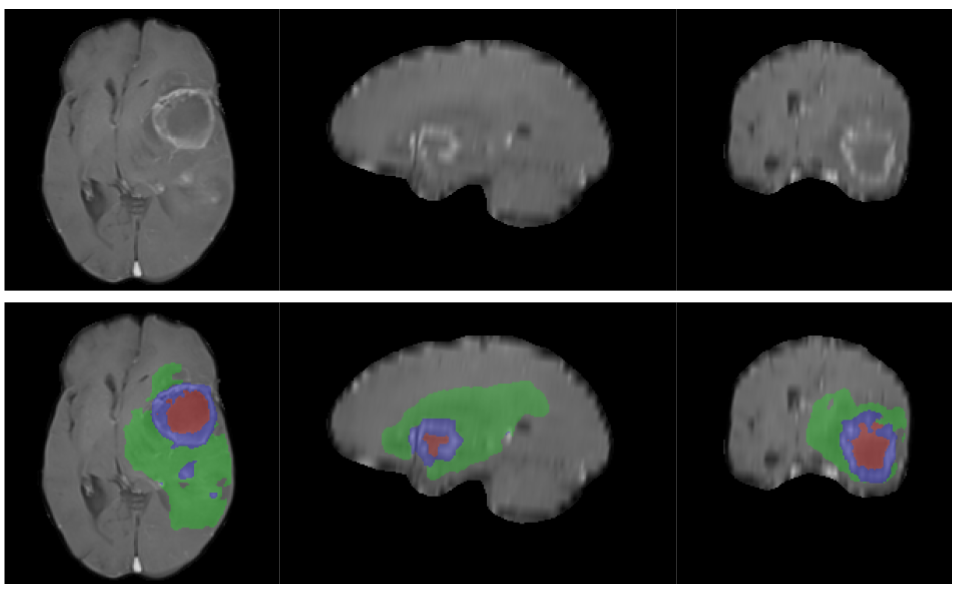}
  \caption{\textbf{Qualitative results} of models after post processing on the validation sample of BraTS-SSA-00227-000. The top row of the figure shows T1CE and the performance of model in segmenting different regions(NCR-red, ED-green, ET-blue).}
  \label{fig:qual-results}
\end{figure}
\subsection{Hyperparameter search}
The first hyperparameter we determined was the ensemble weights. Here, the weights 
$w_n$
  for nnU-Net and 
$w_m$
  for MedNeXt in equation \ref{eqn:1} were chosen based on each model’s performance during cross-validation. These weights were set to 0.4722 for nnU-Net and 0.5278 for MedNeXt, reflecting their performance across the five folds.

The next set of hyperparameters we optimized included the thresholds for adaptive post-processing, as described in section  \ref{sec:adapt}. The results of this search are shown in table \ref{tab:hyper}.

\begin{table}[]
\caption{\textbf{Hyperparameters} estimated on the cross-validated folds for the adaptive post-processing steps.}
\centering
\begin{tabular}{l   c   c   c   c   c   c   c   c   c   }
\hline
\textbf{Cluster} &
  \textbf{0} &
  \textbf{1} &
  \textbf{2} &
  \textbf{3} &
  \multicolumn{1}{l}{\textbf{4}} &
  \multicolumn{1}{l}{\textbf{5}} &
  \multicolumn{1}{l}{\textbf{6}} &
  \multicolumn{1}{l}{\textbf{7}} &
  \multicolumn{1}{l}{\textbf{8}} \\ \hline
\multicolumn{1}{c}{} &
  \multicolumn{9}{c}{\textbf{lesion threshold(voxels)}} \\ \hline
\textbf{\begin{tabular}[c]{@{}l@{}}label 1\\ label 2\\ label 3\end{tabular}} &
  \begin{tabular}[c]{@{}c@{}}0\\ 0\\ 0\end{tabular} &
  \begin{tabular}[c]{@{}c@{}}0\\ 0\\ 50\end{tabular} &
  \begin{tabular}[c]{@{}c@{}}0\\ 0\\ 100\end{tabular} &
  \begin{tabular}[c]{@{}c@{}}0\\ 200\\ 0\end{tabular} &
  \begin{tabular}[c]{@{}c@{}}0\\ 0\\ 50\end{tabular} &
  \begin{tabular}[c]{@{}c@{}}0\\ 50\\ 0\end{tabular} &
  \begin{tabular}[c]{@{}c@{}}0\\ 0\\ 0\end{tabular} &
  \begin{tabular}[c]{@{}c@{}}0\\ 0\\ 0\end{tabular} &
  \begin{tabular}[c]{@{}c@{}}0\\ 0\\ 0\end{tabular} \\ \hline
\multicolumn{1}{c}{} &
  \multicolumn{9}{c}{\textbf{ET/WT ratio threshold}} \\ \hline
 &
  0.0 &
  0.0 &
  0.0 &
  0.0 &
  0.0 &
  0.0 &
  0.1 &
  0.0 &
  0.0 \\ \hline
\end{tabular}%
\label{tab:hyper}
\end{table}

\subsection{Results}

\begin{table}[htbp]
\caption{\textbf{Validation quantitative results} on the SSA dataset using pre-trained (PT) or trained-from-scratch networks. Lesion-wise (LW) Dice coefficients and 95\% Hausdorff distances (HD95) were computed for enhancing tumor (ET), tumor core (TC), and whole tumor (WT), respectively. }
\centering
\begin{tabular}{@{}clllccclccc@{}}
\toprule
\multirow{2}{*}{\textbf{Task}} &  & \multicolumn{1}{c}{\multirow{2}{*}{\textbf{Model}}} & \textbf{} & \multicolumn{3}{c}{\textbf{LW Dice}} & \textbf{} & \multicolumn{3}{c}{\textbf{LW HD95 (mm)}} \\ \cmidrule(l){4-11} 
 &  & \multicolumn{1}{c}{} & \textbf{} & \textbf{ET} & \textbf{TC} & \textbf{WT} & \textbf{} & \textbf{ET} & \textbf{TC} & \textbf{WT} \\ \midrule
\multirow{6}{*}
{\textbf{SSA}} 
&  & nnU-Net (no PT) &  & 0.813 & 0.808 & 0.886 &  & 34.884 & 38.239 & 21.507\\
\cmidrule(l){2-11}
&  & MedNeXt &  & 0.838 & 0.829 & 0.916 &  & 31.409 & 34.709 & 14.160 \\
&  & nnU-Net &  & 0.868 & 0.864 & 0.925 & &20.692 &  23.962 & 14.129 \\
\cmidrule(l){2-11} 
 &  & Ensemble &  & 0.870 & 0.865 & 0.927 &  & 20.668 & 23.949 & 14.003 \\
 &  & Post-processing &  & 0.870 & 0.865 & 0.926 & & 20.745 & 23.950 & 4.003  \\ \bottomrule
\end{tabular}%

\label{tab:val-results}
\end{table}

Table \ref{tab:val-results} provides a comprehensive overview of the performance evaluation of our models across the validation datasets for the SSA task. The challenge's digital platform performed this performance evaluation automatically, without access to the validation ground truth data. Additionally, table \ref{tab:val-results} offers quantitative comparisons of a nnUNet trained from scratch using the SSA dataset to emphasize our proposed approach's importance in increasing performance. Further, Fig.~\ref{fig:qual-results} shows qualitative results in one of the cases. The model achieved the best performance among all participants of the challenge and the test quantitative result. 


\section{Conclusion}
Segmentation of gliomas from radiology images is challenging in resource-limited regions like Sub-Saharan Africa due to scarce and low-quality MRI data. In response, we developed a transfer learning-based approach for adult glioma segmentation that utilizes pre-trained deep-learning models and a stratified fine-tuning strategy. Our method improved segmentation accuracy over trained-from-scratch models, demonstrating the presented technique's potential to bridge the gap in medical imaging capabilities between resource-limited and developed regions. By tailoring machine learning approaches to the specific needs and constraints of the target population, we highlighted the importance of leveraging advanced techniques to enhance diagnostic capabilities in challenging environments. Future work will focus on integrating more local data and refining the stratification process to ensure practical applicability and impact.
\begin{credits}
\subsubsection{\ackname} Partial support for this work was provided by the National Cancer Institute (UG3 CA236536) and by the Spanish  Ministerio de Ciencia e Innovación, the Agencia Estatal de Investigación, NextGenerationEU funds, under grants PDC2022-133865-I00 and PID2022-141493OB-I00, and EUCAIM project co-funded by the European Union (Grant Agreement \#101100633). The authors gratefully acknowledge the Universidad Politécnica de Madrid (www.upm.es) for providing computing resources on the Magerit Supercomputer.

\subsubsection{\discintname}
The authors have no competing interests to declare that are
relevant to the content of this article.
\end{credits}

%
%
%
\bibliographystyle{splncs04}
\bibliography{references}

\end{document}